# Mission to the Gravitational Focus of the Sun:
# A Critical Analysis

**Geoffrey A. Landis**

NASA John Glenn Research Center

21000 Brookpark Road

Cleveland, OH 44135

*geoffrey.landis@nasa.gov*

## Abstract

The gravitational field of the sun will focus light from a distant source to a focal point at a minimal distance of 550 Astronomical Units from the sun. A proposed mission to this gravitational focus could use the sun as a very large lens, allowing (in principle) a large amplification of signal from the target, and a very high magnification. This article discusses some of the difficulties involved in using the sun as such a gravitational telescope for a candidate mission, that of imaging the surface of a previously-detected exoplanet. These difficulties include the pointing and focal length, and associated high magnification; the signal to noise ratio associated with the solar corona, and the focal blur. In addition, a method to calculate the signal gain and magnification is derived using the first-order deflection calculation and classical optics, showing that the gain is finite for an on-axis source of non-zero area.

## 1. Introduction

The fact that the gravity of a massive body deflects light is one of the classical consequences of Einstein's general theory of relativity. This effect means that the sun (or any massive body) can effectively be used as a lens. For the sun, the focal distance of this gravitational lens is well known to be about 550 astronomical units from the sun for the case of light skimming the surface of the sun (the "minimum gravitational focal distance"). The sun continues to act as a lens at any distance beyond this minimum focal distance; at longer distances, the focused light passes increasingly far from the surface of the sun.

Starting with Eshleman in 1979, there have been several suggestions that the gravitational lens of the sun could be used as a telescope. In particular, Maccone has proposed missions to the gravitational focal point. The advantage of this is that the gravitational lens of the sun is, in some sense, a telescope with an area comparable to the size of the sun, far larger than any telescope currently conceived.

This note is to point out the some of the difficulties in practical realization of that concept, and show some of the reasons that a mission to the gravitational lens may be less useful as a telescope than some analysts have suggested.

The challenge of the mission to the necessary distance is not discussed here; proposed methods of reaching the focal distance include high specific-impulse electric propulsion or use of a solar sail. The lure of the mission is that the gravitational focus of the sun is one of very few possible targets for an interstellar precursor mission. Other than this gravitational focus, there is little of any interest at distances between the Kuiper belt, ~50 Astronomical Units, and Alpha Centauri, at about 280,000 astronomical units. There is thus a powerful incentive to find some plausible objective in visiting the gravitational focus, as a potential intermediate step toward a future interstellar mission.

The difficulties with the use of the gravitational lens of the sun as a telescope can be divided into three categories: (1) pointing and focal length, (2) signal to noise ratio, and (3) resolution

The following calculations will discuss as an example case a mission to use the gravitational lens as an optical telescope, but similar considerations apply for other wavelengths.

### *Background*

The deflection of light by the gravity of a massive object is derived in any of the standard textbooks on general relativity (*e.g.,* Schulz 1985). The gravitational deflection angle of light passing a massive body is

$$\theta = (4GM/c^2)(1/r) \tag{1}$$

where r is the distance by which the rays being focused miss the center of the sun. Angles are defined in figure 1 (where the subscript 0 indicates that the light is exactly focused). From this, the distance F to the gravitational focus is:

$$F = (4GM/c^2)^{-1} \, r^2 \qquad\qquad (2)$$

or, expressing r in terms of F:

$$r = (4GM/c^2)^{1/2} \, F^{1/2} \qquad\qquad (3)$$

The condition required to use of the gravitational lens as a telescope is that the light from the distant target reaching the focus must not intersect the surface of the sun, and hence r must be greater than the radius of the sun, about 700,000 km. The case $r = R_{sun}$ is the definition of the minimum gravitational focus.

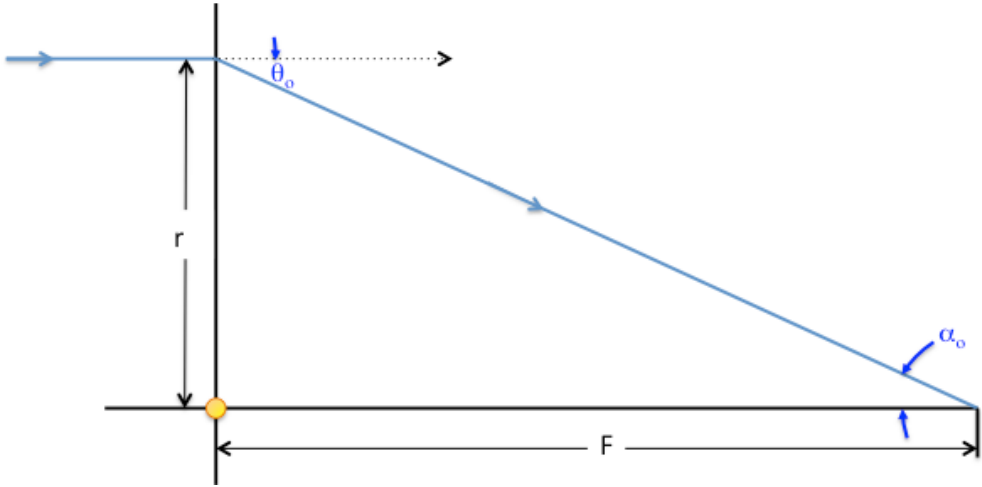

*Figure 1:* definition of angles for gravitational lens

### Einstein Ring

Assuming that $r > R_{sun}$, at the distance F, the rays passing a distance r from the sun form an Einstein ring (as viewed by the observer at F). The angular diameter of the Einstein ring can then be calculated as a function of the distance to the focal point F:

$$\alpha_o = \Theta_o = r/F = (4GM/c^2)^{1/2} \, F^{-1/2} \qquad\qquad (4)$$

At the minimum focal distance of 550 AU, the Einstein ring and the sun both have an angular diameter of 68 microradians, about 2.5 arc seconds.

Figure 2 shows the Einstein ring in schematic, for a distance F that is greater than the gravitational focus distance. In this image, the blue ring is the focused light from the planet. Thus, in this image, the blue ring is the signal that the telescope is observing; everything else in the image can be considered noise.

### Amplification, Gain and Magnification

The most important factor is to calculate the magnification, and hence the gain, of the gravitational lens.

As has been noted in many references (*e.g.,* Bontz and Haugan 1981; Nakamura and Deguchi 1999), the on-axis geometrical gain (or magnification) of a gravitational lens telescope is theoretically infinite. Since infinite gain is clearly not possible in the real world, there is a myth that has been repeated many of the previous analyses that the theoretical "infinite" gain means that the telescope cannot be analyzed using geometrical optics, and thus that a wave optic approach is the only way to calculate the real world performance.[*] This is untrue: in fact, the geometrical optics analysis, in the real world, is always finite, and relatively easy to calculate.

The signal *amplification* is defined here as the flux received at the focus of the optical system, divided by the flux that would be received without the optical system. The *gain* is the amplification expressed in logarithmic units.[†]

*Magnification* is the angular size of the target object as viewed at the focus, divided by the angular size without the optical system. (In this case, we are interested in the *area* magnification, expressed in terms of solid angle, not the linear magnification.)

Because of the brightness theorem (also known as the conservation of étendue), amplification and magnification are the same. (This is a consequence of the more general Liouville theorem, which still holds for the general relativistic case[‡].) For example, a magnifying glass can be used focus the sunlight onto a point, and thus increase the flux of sunlight on the focus. However, viewed from the focal point, the disk of the sun is magnified-- the brightness of any point on the solar disk is constant; the flux is increased because the apparent area of the solar disk is increased.

Thus, we calculate gain by calculating the magnification of the apparent size (solid angle) of the image. The gravitational lens telescope magnifies and distorts the image of the target object. To calculate gain, the distortion is irrelevant; we only need to calculate the total solid angle.

For the on-axis case, the fictional "infinite" magnification is obvious: the mapping of a zero dimensional point source onto a one-dimensional Einstein ring is singular. Since a point is infinitely smaller in area than a one-dimensional line, the magnification is infinite.

However, the flux from a point is irrelevant; real world targets have finite areas. The infinity can be removed simply by integrating over the angular extent of the source.

*Calculation of Magnification*

The planet can be considered as being smeared out into an Einstein ring surrounding the sun, as shown in figure 2. The magnification is the (solid angle) area of

---

[*] As an example, Bontz and Haugan state "A general result of these studies is that there exists certain region (caustics) in which geometric optics predicts an infinite increase in apparent luminosity. Of course, the actual increase in apparent luminosity must be finite. The infinite increases predicted by geometric optics are an indication that the geometric optics formalism is not valid in the vicinity of a caustic." Likewise, Nakamura and Deguchi state: "the geometric optics approximation breaks down near the lens mapping singularity (caustic) where this approximation gives an infinite brightness for a point source," and Elster (1980) states "In the incoherent case geometrical optics gives correct results (except for the focal region)".

[†] Gain (dB) = 10 log (flux$_{magnifed}$/flux$_{incident}$)

[‡] More specifically, conservation of brightness holds in the general relativistic case as long as the source and receiver are not at different gravitational potential. In the case where the gravitational potential varies, the Liouville theorem still holds, but density of states must be redefined as a function of the metric.

the annulus (the Einstein Ring), divided by the solid angle of the target (assumed here to be an exoplanet.)

The gravitational lens demagnifies the apparent angular size of the planet in the radial direction, and magnifies it in the circumferential direction. The angular area of the Einstein ring is the width of the ring times the circumference, $\Omega = 2\pi\alpha w$.

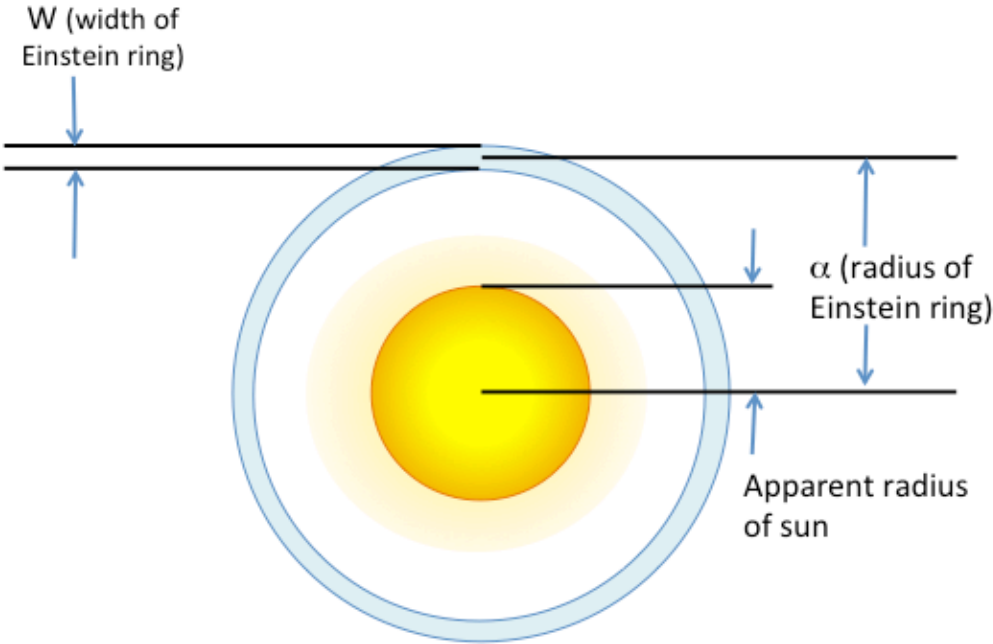

*Figure 2:* View of Einstein ring from focus (not to scale). The blue ring is the focused light from the planet being viewed.

*Radial demagnification*
The radial demagnification is shown in figure 3.

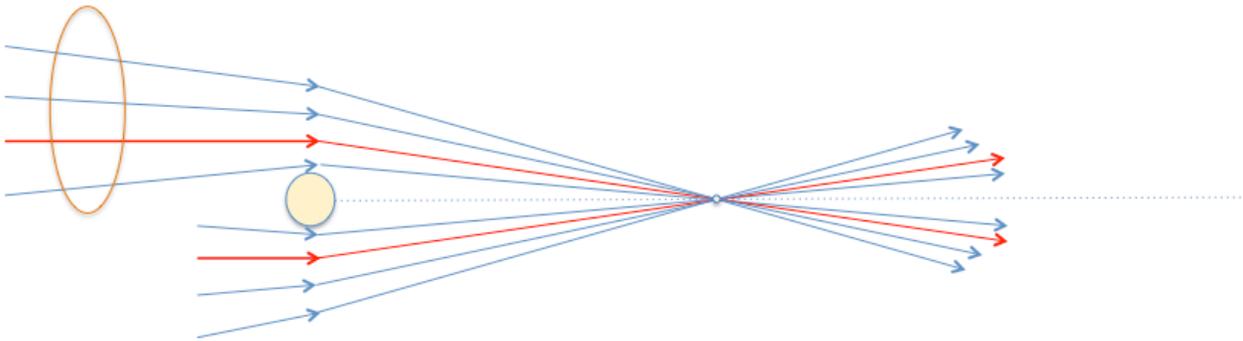

*Figure 3.* Light rays shown in red are the on-axis rays focused at the chosen focal distance. Other rays are out of focus, but still contribute to the light at the focal plane. Note demagnification in the radial direction (rays incident from a large spread of angles reach the focus at a smaller range of angles.) The circled region shows the angular spread of incident rays decreases in the radial direction.

The demagnification factor can be calculated from the geometry, as shown in figure 4. The unfocussed angular diameter of the target planet is $\Delta\theta$, while the focused angular diameter is $\Delta\alpha$. The radial demagnification factor is $\Delta\alpha/\Delta\theta$. Solving this from the gravitational deviation formula (from the previous section), at the gravitational focus the radial demagnification $\Delta\alpha/\Delta\theta$ is exactly 1/2: if the source has an angular diameter a, the apparent width of the source as viewed from the focal point, w, is a/2.

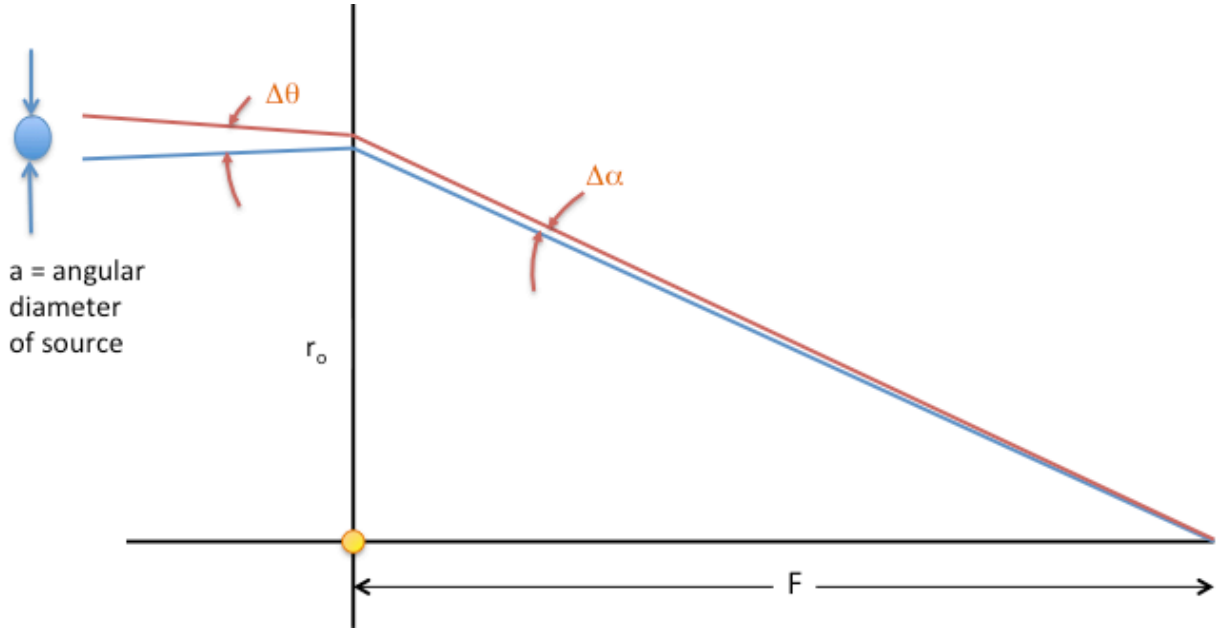

*Figure 4:* radial demagnification factor $\Delta\alpha/\Delta\theta$.

*Total magnification*

With the radial demagnification factor calculated, the total magnification is easily calculated as the area of the annulus divided by the angular area of the source. From figure 1, defining

α = angular radius of Einstein ring

w= angular width of Einstein ring

Ao = (solid angle) area of the annulus = $2\pi r w$

a = angular diameter of source (= $\Delta\theta$ in previous diagram)

Since $\alpha = a/2$

$$Ao = \pi a\alpha \qquad (5)$$

The total area magnification M equals Ao/Ai.

We define Ai as the (angular) area of the source (the exoplanet). For planet of diameter d at distance x, the angular diameter

$$a = d/x \qquad (6)$$

Thus, the angular area of source Ai = $\pi a^2/4$, and the magnification factor is:

$$M = Ao/Ai = (\pi a r)/(\pi a^2/4) = 4\alpha/a \qquad (7)$$

As the angular diameter of the source a approaches zero, the magnification approaches infinity, but the magnification (and hence the amplification) of a real world target of non-zero angular diameter is calculable, and finite.

This calculation can be compared to the result from wave optic calculation. For a target object that is incoherent and not monochromatic, the diffraction fringes average out, and the main consideration of the wave calculation is the resolution of the sensor at the focal plane. In essence, the wave approach calculates the magnification (and amplification) using the assumption that the target size a is the diffraction-limited spot size of the sensor at the focal plane. This results in a $1/\lambda^3$ dependence of amplification on wavelength, where a factor of $1/\lambda^2$ is the gain of the telescope at the focal plane itself, and a $1/\lambda$ factor comes from equation 7, where the spot size a is proportional to wavelength. This calculation, however, can be very misleading, since the target planet is not, in general, the size of the diffraction-limited spot.

From this calculation we can also calculate what fraction of the light at the focal plane comes from what portion of the object being imaged. This is the focal blur, which will be discussed in section 4.

## 2. Pointing and Focal Length

### Pointing

Before calculating the actual performance of the solar gravitational lens as a telescope, it is important to ask whether the telescope would be useful even if it functioned as well as advertised.

A significant difference of the solar gravitational lens from a conventional telescope is that the gravitational lens telescope is not in any practical sense pointable.

For the telescope at a distance F from the sun to be re-aimed to image a new target 1° away, it would have to move a distance of $(\pi/180°)$F, which is 10 astronomical units at the minimum focal distance-- a lateral distance equivalent to the distance from Earth to Saturn. This means that, in practice, such a telescope is not able to be repointed.

Thus, a telescope at the gravitational focus is necessarily going to be a single-purpose telescope, with the target of observation selected before the mission is launched.

A gravitational focus mission can't be used as a telescope to search for a target: such a mission must be with the objective to observe a target whose position is already known.

### Image Size

A related flaw in the concept is the fact that the focal length of the telescope is far too long for the telescope to produce a useable image.

The details depend slightly on what the telescope is intended to do, and how it is designed. I will examine here the case of a telescope designed to image an extraterrestrial planet, and which does so by placing detectors at the focal plane of the gravitational lens.

The geometry is shown in figure 5.

From this figure, it is clear that the size of the object being viewed on the image plane, $X_i$, is related to the size of the object being imaged, $X_o$, by the simple relationship

$$X_i = X_o \, (F/d) \qquad\qquad (8)$$

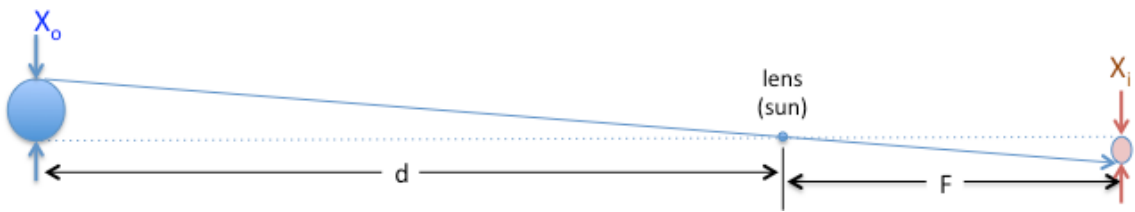

*Figure 5:* geometry of gravitational lens (not to scale).

For the example, to make round numbers, I will use a reference case of a focal plane at 630 astronomical units, which is conveniently equal to 0.01 light years (LY); and assume that the target is a possible planet around a nearby star, taken to be at a distance of 10 LY, approximately the distance of Epsilon Eridani. For the longer focal distance of 2200 AU discussed below, the example distance would be 35 LY.

For this case the size of the image produced at the focal plane can be easily calculated from geometrical optics: the image is smaller than the object by the ratio of F/d (the focal distance divided by the distance to the object), which for the example case is a factor of 1000. If the exoplanet imaged is the diameter of the Earth, 12500 km, the image at the focal plane will be 12.5 km in diameter.

I can't picture any simple way to make a focal plane array many kilometers in diameter. The result is that the telescope does not image the planet, but instead a small fraction of the planet; i.e. for the example case, a focal plane detector one meter in dimension will image a 1-km area on the surface of the planet, a focal plane detector ten meters in dimension will image a 10-km area on the surface, *etc.*. (As noted in section 4, focal blur will make the actual area imaged much larger). One way of mapping the planet might be to raster such an imager across the planetary focal plane.

This brings up a question of pointing: at such an enormous magnification, even if we know that a planet is there, can we even find it with the telescope? This is a telescope that, by its nature, cannot incorporate a conventional "finder scope" at lower magnification-- the target is invisible *except* through the gravity lens. Finding a planet of diameter ~$10^4$ km at a distance of $10^{14}$ km requires a pointing knowledge and pointing accuracy, of 0.1 nanoradians.

Once found, the planet will not stay in the field of view for very long. If the planet is orbiting at the same orbital velocity as the Earth, 30 km/sec, the 1-km section of planet being imaged will traverse a 1-meter focal plane in 33 milliseconds. Rather than imaging the location on the planet, the spot will be motion blurred.

The entire planet will pass across the focal plane in 42 seconds.

This does bring up the possibility of an alternate method of imaging the planet: accepting that the focal plane will not produce an image, but instead allowing the motion of the planet to move across the focal plane and produce a line-scan of brightness (and spectrum) across the surface. The telescope can then be moved to reacquire the planet, and the line scan repeated to slowly build up a raster scan of the surface.

Alternatively, with sufficiently good knowledge of the target planet's orbit, the telescope could move to track the planet. Since an orbital velocity of 30 km/s corresponds to a velocity of 30 m/s at the image plane, the $\Delta V$ to reacquire the planet will be about 30 m/s. Over the course of a year, about 200 m/s would be required to track the planet, with the telescope moving around at the focal plane in an ellipse with a semimajor axis of ~150,000 km. This could be done using some form of high specific-impulse propulsion.

## 3. Signal to Noise Ratio

### *Occulter*

From figure 2, an obvious difficulty is visible: in order to use the gravitational lens as a telescope, the observer has to point directly at the sun. In general, when you read the instructions of what not to do with a telescope, number one on the list will be "avoid pointing directly at the sun." This remains true even at 550 AU, where the sun is still the brightest (and also the most radio-noisy) object in the sky.

To use the gravitational focus as a lens requires that the object being viewed be directly behind the center of the sun, as viewed from the focal observer. The light from the object then forms a ring centered on the sun. At the minimum gravitational focus, this ring exactly touches the visible surface of the sun, but will be separated further away from the sun as the focal distance increases beyond the minimum.

The signal from the sun is far brighter than the gravitationally-lensed image (the Einstein ring), so some sort of occulter or coronagraph has to be used. This is usually handwaved away, but the problem is not trivial. The added difficulty is that the sun cannot be considered a point source, since the angular diameter of the solar disk is comparable to (although necessarily smaller than) the separation between the disk and the Einstein ring to be imaged.

The problem is similar to that analyzed for use of occulters or coronagraphs to directly image extrasolar planets, and there exists considerable literature on that subject which can be applicable here, so other than pointing out that the problem must be addressed, I will not go into further details here.

The light from the parent star will also need to be blocked. Despite the amplification of the gravitational lens, the parent star will still be many orders of magnitude brighter than the planet, and will be only a small distance (under an arc second) away from the planet.

Other stray light must be minimized. The telescope will be looking into the zodiacal light, and the cumulative contribution from stray starlight on the focal plane will also be a source of background, and needs to be minimized. With conventional telescopes, this is done with a collimating tube.

### Solar Corona

The Einstein ring being imaged will overlap the corona of the sun. It is a useful calculation to evaluate how far from the sun the ring needs to be for the Einstein ring of the object to be visible against the brightness of the corona.

Since, per equation (4), the angular diameter of the Einstein ring shrinks as the inverse square root of distance, while the angular diameter of the sun shrinks inversely proportional to distance, the ring gets proportionally further away from the sun as distance increases, even though the absolute angular distance of the ring from the sun decreases.

Many references show the brightness of the solar corona as a function of distance from the solar disk [November and Koutchmy 1996], [Kimura and Mann 1998], [Leinert 1998], [Lang 2010], [Kramer 2002]. The coronal brightness consists of two parts, the light emitted from the corona itself (the "K corona"), and the light scattered from dust (the F corona, which merges smoothly into the zodiacal light). For our purposes, both of these contribute to the background brightness.

November and Koutchmy 1996 fit the brightness as a function of angular distance from the sun to an empirical formula:

$$\frac{I(R)}{I_0} = 10^{-6} \left( \frac{3{,}670}{R^{18}} + \frac{1{,}939}{R^{7,8}} + \frac{0{,}0551}{R^{2,5}} \right) \; ; \quad R > 1 \; , \tag{9}$$

where: $I_0$ is the intensity in the middle of the apparent solar disc, and $R$ the angular distance from the sun's center in units of the angular radius of the sun's disc.

Figure 7 (from Lang 2010) shows that the brightness of the solar corona equals that of the night sky illuminated by the full moon at a distance of 2 solar radii from the center of the sun (that is, one solar radius from the edge). Since astronomical observations of faint objects are not taken during periods of full moon due to the sky brightness, it seem reasonable to expect that imaging will require a background less bright than this, so 2 solar radii can be taken as a reasonable distance of the Einstein ring from the sun for the gravity-lens telescope to be useful. (As will be seen later, in the discussion of signal to noise ratio, this value in actual operation must be significantly larger.

(it is worth pointing out that, due to Liouville's theorem, the surface brightness of the object being observed is independent of the lensing. The gravitational lens will increase the total light received from object, but this is because the focusing magnifies the size of the image, increases the observed surface area. The surface itself is of constant brightness.)

From equation 4, the angular separation of the disk from the sun is twice the apparent solar radius at a distance of 4 times the minimum focal distance, or 2200 AU.

(This is likely to be an underestimate. The light from the corona of the sun extends much further out than the Einstein ring, and it is the integrated light from the corona that is masking the signal.)

Thus, when the corona of the sun is considered, the useful gravitational focus of the sun is much more distant: at least 2200 AU, not the 550 AU usually quoted.

The solar corona also will have a refractive effect on radio waves. A discussion this can be found in Turyshev and Anderson 2003.

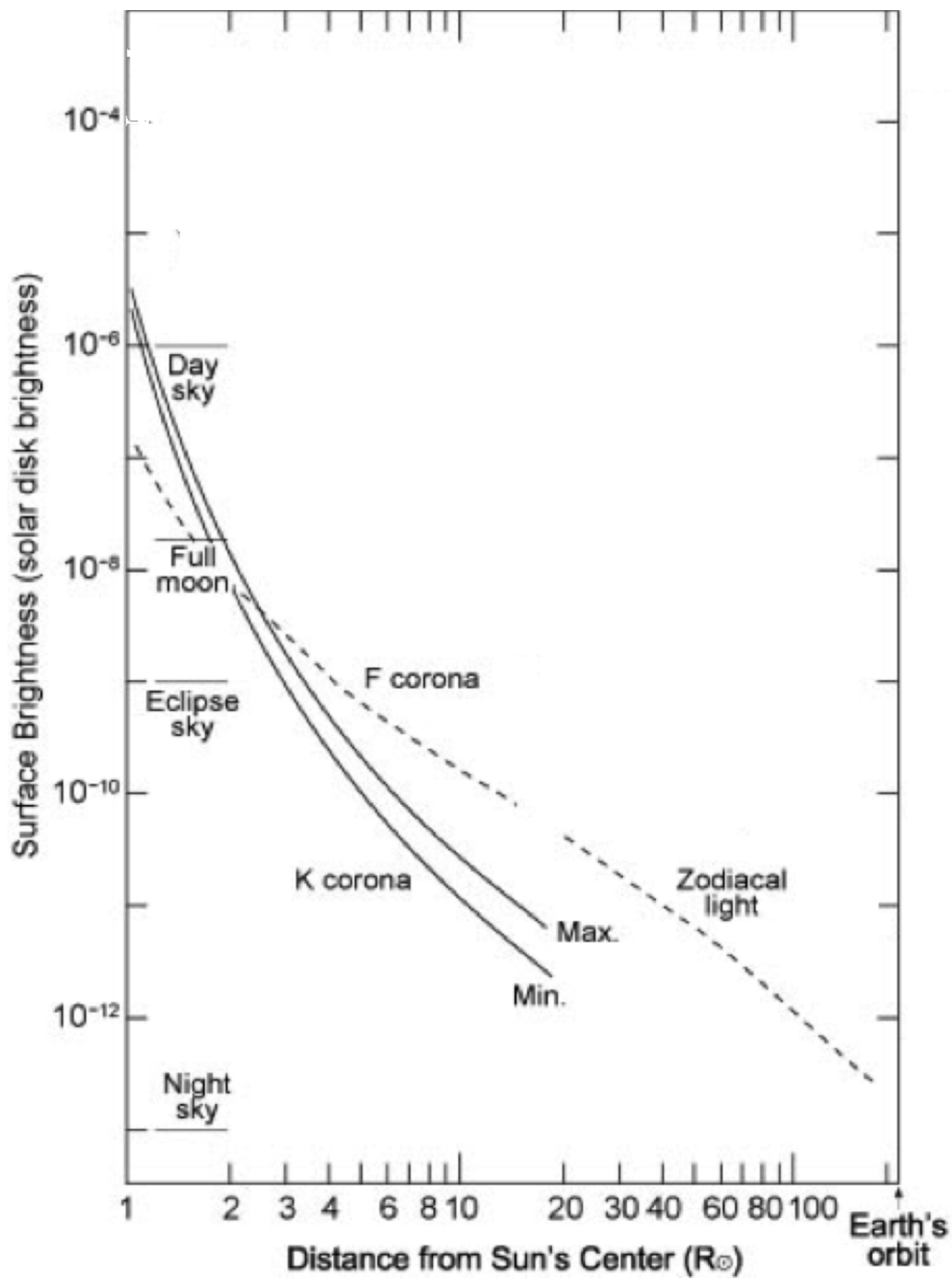

*Figure 7:* Brightness of solar corona compared to the sky brightness during the day, a moon-illuminated night, during a solar eclipse, and the unilluminated night sky. From Lang 2010.

### Target Star

In addition to the solar brightness, the parent star of the exoplanet being observed is also a source of noise in the system. The amount of separation of the planet from the star will vary depending on where the planet is in its orbit, but at the most favorable position, we can expect the separation of an earthlike planet from the parent star to be about half a second of arc for a planet exactly one parsec away, varying inversely with distance. The fact that the star can be brighter than the planet by a factor on the order of 1E10 is the reason that planets are not directly imaged by ground-based telescopes; this large difference in brightness means that even though the stars brightness is not amplified by the gravitational lens, it will nevertheless be a significant source of noise unless its light also blocked by a coronagraph.

### Signal Amplitude: summary

Discussions of gravity lens telescopes tend to emphasize the high magnification of the telescope. Nevertheless, the signal source is still only 3.5 arc seconds across, so even with the high magnification (and consequent intensity enhancement), the absolute intensity of the signal is still small.

A calculation of the magnification using wave optics can be found in several sources, such as Turyshev and Andersson 2003. However, this method of calculation necessarily incorporates both the optics of the gravitational lens itself as well as the optics at the focal plane (since the achievable magnification is assumed to be limited by diffraction at the focal plane). Rather than use a wave-optics calculation, it is useful to calculate the geometrical optics, and then analyze the diffraction separately. For our purposes, it is relatively simple to calculate the total flux from the planet being imaged.

At a distance of 10 LY, a planet of 10,000 km diameter subtends 1E-10 radians. The gravitational lens demagnifies the image of planet in the radial direction by exactly a factor of two. Since the subtended radius of the ring in our example case is 15 microradians, the total area of the ring, which will the total area of sky subtended by the magnified planet, is ~3 E-15 steradians. If we assume planetary albedo typical of Earth, the amount of energy reflected will be about 400 watts/square meter. The total energy flux from the planet received at the focal plane is thus about E-12 watts per square meter. Although it sounds like a small number, this corresponds to an intensity increase of nearly 100,000 over the unlensed planet.

However, this factor of 100,000 is against the bright solar corona, discussed above (as well as the diffraction tail of solar brightness getting past the coronagraph). We want the signal to be greater than the noise.

The maximum signal to noise ratio comes if we block all light except that from the Einstein ring. However, this would require resolving the Einstein ring, and if we could do that, since the ring has an angular diameter half that of the planet, we could just resolve the planet directly

But this still brings up a critical question. Given all the difficulties discussed above, is it worth travelling out to beyond 600 AU to merely gain a factor of 100,000? Is this enough?

## 4. Focal Blur

The magnification and gain of the gravitational lens were calculated in section 1. From this, we can easily derive the focal blur.

Since the flux from an angular source of a given brightness is proportional to the angular diameter squared, while the magnification is inversely proportional to the angular diameter, the total flux received from the object decreases toward zero as the object gets smaller (even though the magnification approaches infinity, in this case infinite magnification of an infinitesimal object is infinitesimal. This is because the area approaches zero faster than the magnification approaches infinity.)

From the equation 7 in section 1 it is now possible to calculate the focal blur. Although the central spot on the target planet is intensified relative to the rest of the planet (geometrical magnification of central spot is infinite... but only for an infinitesimal area); most of the light received at the focal plane is not from the central spot. Focal blur is inherent in the gravitational lens; it does not change with position or magnification.

Figure 8 shows a candidate planet being imaged. Note that 50% of the light reaching the focal point comes from the inside the 50% diameter; and 50% comes from outside the 50% diameter (regardless of the fact that the inner 50% of the diameter comes to only 25% of the area). Likewise, 10% of the light reaching the detector comes from the central 10% of the diameter, accounting for 1% of the planet's area. Although the central regions are weighed more heavily in the amplification (gain), the outer parts have proportionally larger area. Thus, if we define the FWHM focal blur as the circle inside which half the light originates, *the focal blur is exactly half the diameter of the planet, regardless of the size of the planet.*

Correcting the focal blur could be done if the telescope at the focus was able to resolve the width of the Einstein ring. But because of the radial demagnification of the gravitational lens, the width of the Einstein ring is half the angular width of the planet, and hence any telescope that could resolve the width of the Einstein ring could image the planet directly, without need for the gravitational lens.

With sufficiently sophisticated deconvolution technique, it may be possible to sharpen the image, using the fact that the portion of the image closer to the axis contributes proportionally more to the total image, and also possibly taking the rotation of the planet into account. However, it is clear that the magnification of the planetary image at kilometer scales cannot be achieved.

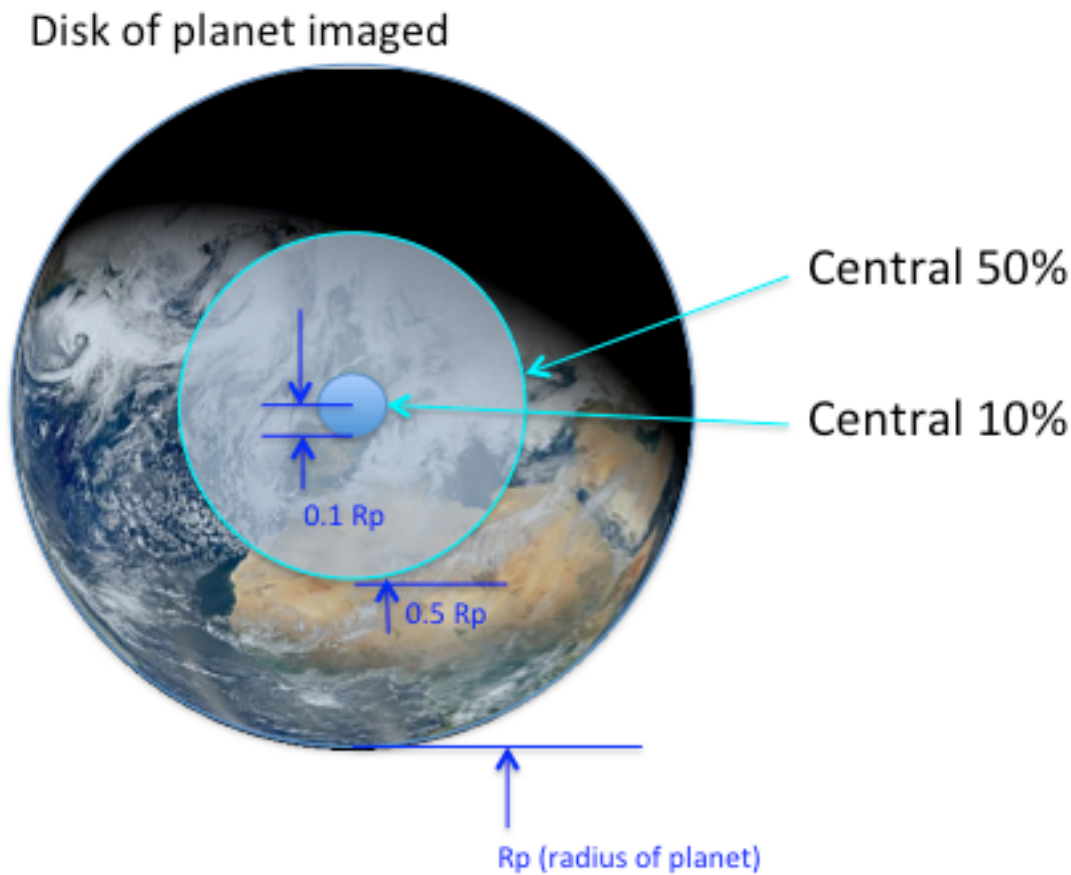

*Figure 8:* focal blur of the image of a candidate planet

### *Imaging the Einstein Ring*

It was earlier stated that the surface of the planet was smeared out into the Einstein ring. More precisely, the gravitational lens maps the surface of the planet onto the Einstein ring. It is interesting to look in more detail at this mapping.

The point exactly on the optical axis is mapped to the central circle of the Einstein ring. Every other point on the planet is mapped onto the Einstein ring twice, in mirror imaged once inside and once outside of the central circle. This is shown in schematic in figure 9, for the case where the planet is not centered on the optical axis, but is displaced slightly to the right.

The width of the Einstein ring, of course, is far too narrow to be resolved by a telescope at the focal plane. However, it takes only a relatively modest telescope to resolve the circumference of the ring, 2.5 arc seconds at the closest focal distance. Each point on the circumference of the Einstein ring averages the light received from a stripe across the planet's disk (with each stripe repeated twice at positions 180° around the disk). It is possible to think that the planet's area can be reconstructed by these stripes.

At any given position of the telescope at the focus of the gravitational lens, there is an mirror ambiguity in that there is no way to distinguish light from the left end of the stripe from the right end. This ambiguity may be resolved by views from more than one location of telescope at the gravitational focus, or by allowing the moving image of the planet to pass across the telescope, thus changing the center spot.

Since the planet would be rotating, additional information can be retrieved by the fact that surface features will slowly rotate into and then out of view, and from the fact that part of the planet is illuminated and part is dark. Multiple images of the planet may be required to distinguish the changes due to planetary rotation from the changing cloud patterns on the planet.

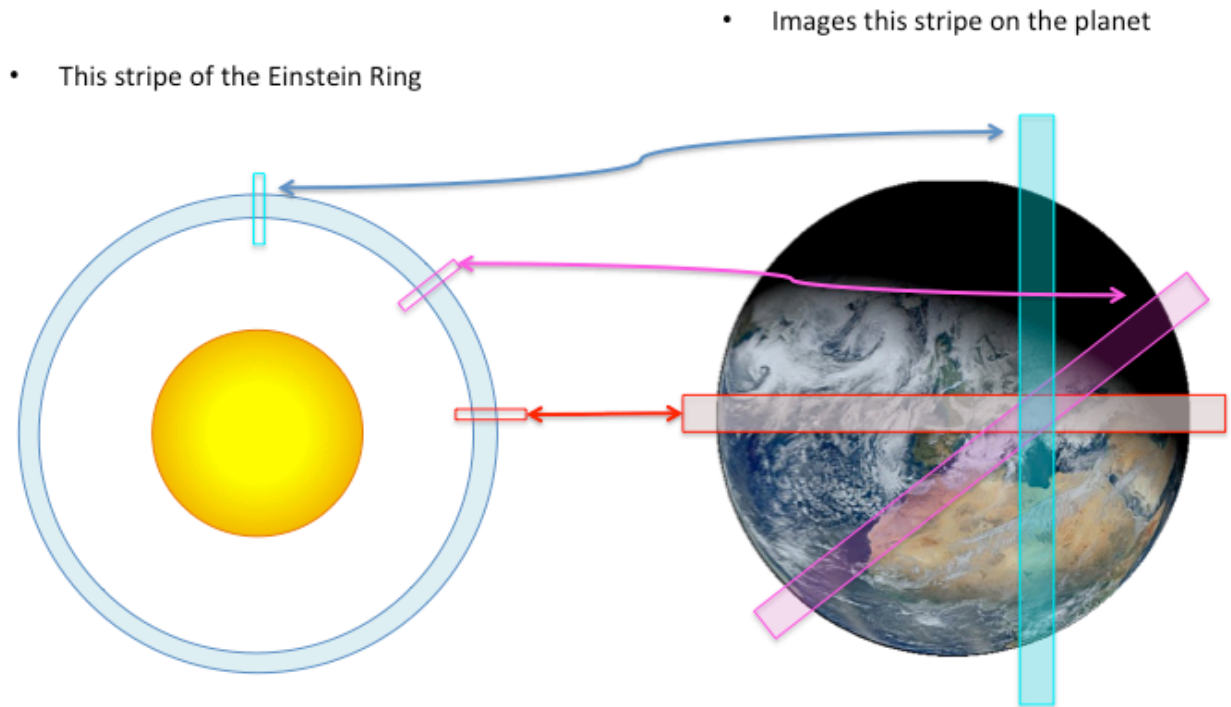

*Figure 9*: Mapping of the planet to the Einstein ring. In this example case, the planet is not centered on the optical axis, but is displaced slightly to the right

---

[§] See figure 2 "Solar radial log-intensity variation", and equation 1.

[**] See figure 57: "The visible equatorial and polar F-corona brightnesses in comparison to typical values for K-corona, the aureole (circumsolar sky brightness enhancement) and instrumental stray light levels"

[††] See Figure 8: "The brightness profiles, in units of mean solar brightness, of the unvignetted K-corona, F-corona, and the MiniCOR maximum expected instrumental background over the 2.5 R to 20 Rs field of view".